\documentclass[12pt]{iopart}

\usepackage{graphicx}
\usepackage{amsfonts}

\usepackage{epsfig}
\usepackage{graphicx}% Include figure files
\usepackage{dcolumn}% Align table columns on decimal point
\usepackage{bm}% bold math
\usepackage{xcolor}

\begin{document}
    
    \title{Effective edge-based approach for promoting the spreading of SIR model}
    
    \author{Dan Yang$^{1}$, Jiajun Xian$^{2*}$, Liming Pan$^{3,2}$, Wei Wang$^{4,2}$, Tao Zhou$^{1}$}
    
    \address{$^{1}$ Web Sciences Center, School of Computer Science and Engineering,
        University of Electronic Science and Technology of China, Chengdu, 611731, China}
    
    \address{$^{2}$ School of Computer Science and Engineering,
        University of Electronic Science and Technology of China, Chengdu, 611731, China}
    
    \address{$^{3}$ School of Computer Science and Technology, Nanjing Normal University, Nanjing, Jiangsu, 210023, China}
    
    \address{$^{4}$ Cybersecurity Research Institute, Sichuan University, Chengdu, 610065, China}
    \address{E-mail: *xianjiajun22@gmail.com}

    %\ead{custserv@iop.org}
    
    \begin{abstract}    
 Promoting some typical spreading dynamics, for instance, the spreading of information, commercial message, vaccination guidance, innovation, and political movement, can bring benefits to all aspects of the socio--economic systems. In this study, we propose a strategy for promoting the spreading of the susceptible--infected--recovered model, which is widely applied to describe these common spreading dynamics in real life. Specifically, we first quantify the potential influence that the addition of each latent edge (that is, edges that do not exist before) could cause to the spreading dynamics. Then, we strategically add the latent edges to the original networks according to the potential influence of each latent edge. Numerical simulations verify the effectiveness of our strategy and demonstrate that our strategy outperforms several static strategies, namely, adding the latent edges between nodes with the largest degree or eigenvector centrality. This study provides an effective way of promoting the spreading of the susceptible--infected--recovered model by modifying the network structure slightly and helps in understanding what a better network structure for the spreading dynamics is. Besides, the theoretical framework established in this study provides inspirations for the further investigations of edge--based promoting strategies for other spreading models.      
    \end{abstract}
    %89.75.Hc: Genealogical trees (complex systems)
    %87.19.X-: Disease
    %64.60.Ht: dynamic critical behavior

    %Uncomment for PACS numbers title message
    %\pacs{00.00, 20.00, 42.10}
    % Keywords required only for MST, PB, PMB, PM, JOA, JOB?
    %\vspace{2pc}
    %\noindent{\it Keywords}: Article preparation, IOP journals
    % Uncomment for Submitted to journal title message
    %\submitto{\JPA}
    % Comment out if separate title page not required
    
    \maketitle
    \tableofcontents
    \section{Introduction}

The subject of promoting the spreading dynamics in networked systems is attracting substantial attention from multiple disciplines, for instance, computer science, statistical physics, and network science~\cite{morone2015influence,pan2019optimal}. Maximizing the spreading prevalence of some common spreading dynamics, including the spreading of information, vaccination guidance, innovation, commercial message, and political movement, can bring benefits to all aspects of the socio-economic systems~\cite{wang2016statistical,gong2016influence,del2018finding,hu2018local,lokhov2017optimal}. The study of promoting these spreading dynamics is of great importance in both theoretical and practical perspectives.     
    
Understanding the evolutionary mechanisms of the spreading dynamics in real life and building suitable models to describe them play essential roles in developing promoting strategies. 
Various spreading models have been proposed for spreading dynamics with different evolutionary mechanisms. 
For instance, in some simple contagions (e.g., information diffusion and innovation spreading) where the infected individuals could infect the susceptible ones by a single contact, the classic susceptible--infected--susceptible (SIS) model~\cite{fu2008epidemic,xian2020optimal}, susceptible--infected--recovered (SIR) model ~\cite{daley1964epidemics,daley1965stochastic} and many of their extensions~\cite{wang2014multiple, sun2016pattern,zuzek2015epidemic,wang2019impact,xia2019new} have been widely applied. Besides, for some complex contagions (e.g., behavior adoption~\cite{aral2017exercise}  and political information spreading~\cite{romero2011differences,centola2007complex}), researchers have proposed the threshold model which incorporates the social reinforcement mechanism (i.e., the mechanism that the susceptible individuals becoming infected with a probability that increases with the cumulative number of contacts with the infected ones)~\cite{watts2002simple,lu2011small}. More spreading models with other complex mechanisms can be found in ~\cite{wang2019coevolution,boccaletti2014structure,wang2015coupled}.

Based on these spreading models, researchers go further to develop strategies to promote or enhance the spreading dynamics. Some of the researchers focus on designing effective transmission strategies~\cite{yang2008optimal,gao2016effective,yang2008selectivity,roshani2012effects,gao2017promoting,cui2018close} such as developing smart protocols to avoid invalid contacts (for instance, the contact between two infected nodes). Besides, some of the researchers are committed to identifying vital nodes~\cite{lu2016vital,liao2017ranking,xin2019discerning,chen2010scalable,riquelme2016measuring,Kitsak2010Identifying,lu2016h,chen2009efficient} with high centralities (for instance, degree, betweenness, and closeness centrality); and they suggest that selecting these vital nodes to the initial seeds can maximizing the spreading prevalence. Recently,  several researchers find that structural perturbations (that is, modifying the network structure slightly) can be used for promoting spreading dynamics as well~\cite{aguirre2013successful,milanese2010approximating,van2010influence}.  

Nevertheless, despite all of these efforts, no previous study has investigated the problem of how to effectively promote the spreading dynamics of the SIR model by structural perturbations, to the best of our knowledge. To fill up this research blankness,  we propose an effective edge--based strategy for promoting the SIR spreading dynamics in this study. The SIR model is first proposed to study the epidemic transmission. Later on, researchers extend it to various other contagion processes, including information diffusion, innovation spreading, promotion of commercial products and the spread of political movements~\cite{xian2019misinformation,zhu2014information,shulgin1998pulse,zhao2013sir,chen2016detecting,wu2004simulation,woo2011sir,xian2020containing}. Our strategy enhances the spreading dynamics of the SIR model by adding edges that do not exist before.
To be specific, we first develop a mathematical model to quantify the influence that the addition of each latent edge (i.e., each edge that does not exist in the original network) could cause to the spreading dynamics. This developed mathematical model is able to facilitate the determination of the spreading prevalence of the SIR model as well.
Then, we strategically add the latent edges to the original networks according to the influence of each latent edge. Note that our strategy incorporates both the information of network structure and spreading dynamics. This study will show that our strategy is effective and outperforms those static approaches, such as adding the latent edge between nodes with the highest degree or eigenvector centrality.

We organize this paper as follows. First, Sec. \ref{sec:model} describes the spreading model and our strategy in detail. Then, Sec. \ref{sec:theory} gives the theoretical framework for determining the influence of each latent edge. Further,  Sec. \ref{sec:simulation} presents the numerical simulations to verify the effectiveness of our strategy. Finally, Sec. \ref {sec:conclusion} concludes the paper.

\section{Model description} \label{sec:model}

In this study, we consider a discrete-time SIR dynamics that runs on a complex network $G$ with adjacency matrix $A$. The number of nodes and edges of $G$ is denoted by $N$ and $M$, respectively. Generally, each node in this model will be assigned with one of three different states, that is, the susceptible state (S), the informed (or infected) state (I), or the recovered state (R). Denote the state of node $i$ by $\varepsilon_i$; thus, $\varepsilon_i \in \{S, I, R\}$. 
    Initially, all the nodes are set to be in the S state. Then, a small fraction of nodes are selected to be in the I state. For every time step, every node in the I state will infect or inform each of its neighbors in the S state with the transmission probability $\lambda$.
    After the transmission process, each node in the I state will turn to the R state with the recovery probability $\gamma$. We refer to $\beta=\lambda/\gamma$ as the effective transmission probability.
 The spreading dynamics will be terminated once there is no node in the I state, and the fraction  $\rho$ of nodes in the R state after the termination of spreading dynamics is referred to as the spreading prevalence.

According to the evolutionary rules of the SIR model described in the above paragraph, we can obtain the probabilities of nodes and edges in different states when the dynamics is terminated, for instance, the probability $P(\varepsilon_i = R)$ of node $i$ being in R state or the joint probability $P(\varepsilon_i = R, \varepsilon_j = S)$ of edge $(i,j)$ being in RS state.
    Our objective is to maximize the spreading prevalence of the discrete-time SIR dynamics that runs on top of the network $G$ by adding a fraction of latent edges, i.e., the edges that do not exist in the original network $G$ before. 
    To determine which latent edge should be added first, we need a measure to rank the influence of each latent edge.
    
 Consider we add a latent edge $(i,j)$ to the original network $G$. If the final states of nodes $i$ and $j$ are $\varepsilon_i =\varepsilon_j= S$, then this added edge will make no difference to the spreading prevalence since both node $i$ and $j$ will still be in the S state and influence no other node. 
    Similarly, if the final states of nodes $i$ and $j$ are supposed to be $\varepsilon_i =\varepsilon_j= R$, then adding an edge between them will barely bring new nodes to the I state because nodes $i$ and $j$ will be infected or informed regardless of whether they are directly connected.
    Therefore, only when the final states are $\varepsilon_i =R$ and $\varepsilon_j= S$ (or $\varepsilon_i =S$ and $\varepsilon_j= R$), the spreading prevalence will be increased by adding an edge between nodes $i$ and $j$. Take the former situation as an example, that is, the situation when the final states of nodes $i$ and $j$ are $\varepsilon_i =R$ and $\varepsilon_j= S$, respectively. 
    In this case, if we add an edge between nodes $i$ and $j$, and node $i$ gets infected or informed in the time $t_{0}$, then node $i$ can bring node $j$ into the I state with probability $\lambda$ in the time $t_{0}+1$. 
 When it comes to the time $t_{0}+2$, as a new node in the I state, node $j$ goes ahead to influence its neighbors in the S state. 
    Obviously, if node $j$ has a large expected number of neighbors whose final states are S, then adding the edge $(i,j)$ can bring a large number of new nodes into the I state and increases the final spreading prevalence. 
Therefore, we only consider node $j$ and its neighbors whose final states are S. For convenience, we refer to node $j$ and its neighbors who have a final state of S as the candidate nodes. 
Then, the expected number of new infected or informed nodes that come from the candidate nodes after adding the latent edge $(i,j)$ can be calculated as
    \begin{equation}\label{eq:nij}
    {\overline \sigma}_{ij}=\lambda P(\varepsilon_i =R)P(\varepsilon_j =S)[1+\sum_{r=1}^NA_{jr}\lambda P(\varepsilon_r=S\vert\varepsilon_j=S)],
    \end{equation}
    where $P(\varepsilon_r=S\vert\varepsilon_j=S)$ is the conditional probability that node $r$ is in the S state when $j$ is in the S state. 
    Similarly, we can obtain the expected number ${\overline \sigma}_{ji}$ when the final states of nodes $i$ and $j$ are $\varepsilon_i =S$ and $\varepsilon_j= R$, respectively.
    Take both cases of ${\overline \sigma}_{ij}$ and ${\overline \sigma}_{ji}$ into consideration, we define the influence of latent edge $(i,j)$ as
    \begin{equation}\label{eq:Iij}
        \sigma_{ij}={\overline \sigma}_{ij}+{\overline \sigma}_{ji}.
    \end{equation}
    Our approach to effectively promote the spreading dynamics of the SIR model is based on adding the latent edge with the highest influence $\sigma_{ij}$. Thus, we refer to our strategy as the latent--edge--influence (LEI) strategy. 
    Hereafter, the problem reduces to solving Eq. (\ref{eq:Iij}), that is, finding the probabilities of nodes in different states and the conditional probabilities.

\section{Theoretical analysis} \label{sec:theory}

    In this section, we will develop a new theoretical framework to study the discrete--time SIR spreading dynamics on complex networks.  Based on this developed framework, Eq. (\ref{eq:Iij}) can be well solved.
    
    Inspired by the epidemic link equations (ELE) model proposed by Matamalas et al.~\cite{matamalas2018effective}, we first define a set of discrete--time equations for the probabilities of edges in different states and then solve the equations at the final state.    For the sake of simplicity, we denote the joint probabilities $P(\varepsilon_i = X, \varepsilon_j = Y)$ as $\Theta^{XY}_{ij}$, where $X,Y \in \{S, I, R\}$. The evolution of these denoted joint probabilities depends on each other according to the evolutionary rules of the SIR model.
    
    For instance, the iteration of  $\Theta^{II}_{ij}(t)$  sponges on  $\Theta^{SS}_{ij}$, $\Theta^{SI}_{ij}$,  and $\Theta^{IS}_{ij}$. Specifically, we can obtain  the iteration formula of $\Theta^{II}_{ij}(t)$ as follows:
        \begin{eqnarray}    \label{II}             
    \Theta^{II}_{ij}(t+1)&=&\Theta^{SS}_{ij}(t)(1-q_{ij}(t))(1-q_{ji}(t)) \nonumber \\
    &+&\Theta^{SI}_{ji}(t)(1-\gamma)(1-(1-\lambda)q_{ji}(t)) \nonumber \\
        &+&\Theta^{SI}_{ij}(t)(1-\gamma)(1-(1-\lambda)q_{ij}(t))+\Theta^{II}_{ij}(t)(1-\gamma)^{2},
    \end{eqnarray}
where $q_{ij}(t)$  represents the probability that node $i$ (in the S sate) is not brought into the I state by any of its neighbors (excluding $j$).  Note that Eq. (\ref{II}) has taken into account all the possible state changes of nodes $i$ and $j$.  Given the states of nodes $i$ and $j$ at time $t+1$ as $\varepsilon_i(t+1)=\varepsilon_j(t+1)=I$, the first term of Eq. (\ref{II}) considers the situation when  $\varepsilon_i(t)=\varepsilon_j(t)=S$ and both nodes $i$ and $j$ are brought into the state I by their neighbors at time $t$. Besides, the second term represents that the states of nodes $i$ and $j$ at time $t$ are $\varepsilon_i(t)=I$ and $\varepsilon_j(t)=S$, respectively, and  then node $i$ holds its state but node $j$  is brought into the state $I$ by its neighbors. Moreover, the third term accounts for that the state of node $i$ $(j)$ is $\varepsilon_i(t)=S$ $[\varepsilon_j(t)=I]$ at time $t$ and then node $i$ is brought into the state $I$ while node $j$ holds its state. Last, the fourth term considers that nodes $i$ and $j$ are both in the state $I$ at time $t$ and remain in the state $I$ when it comes to time $t+1$.
    
    Similarly, the iteration formulas of joint probabilities $\Theta^{SS}_{ij}(t)$ and $\Theta^{RR}_{ij}(t)$ can be obtained as 
    \begin{eqnarray}    
    \Theta^{SS}_{ij}(t+1)&=&\Theta^{SS}_{ij}(t)q_{ij}(t)q_{ji}(t)
    \end{eqnarray}
and
    \begin{equation}    
    \Theta^{RR}_{ij}(t+1)=\Theta^{II}_{ij}(t)\gamma^{2}+\Theta^{RI}_{ij}(t)\gamma+\Theta^{RI}_{ji}(t)\gamma+\Theta^{RR}_{ij}(t),
    \end{equation}
respectively.    Note that for the joint probability $\Theta^{XY}_{ij}(t)$, where $X=Y$ and $X \in \{S,I,R\}$, we should have $\Theta^{XY}_{ij}(t)=\Theta^{XY}_{ji}(t)$. However, if $X\neq Y$, $\Theta^{XY}_{ij}(t)$ and $\Theta^{XY}_{ji}(t)$ may have different values. That is to say, we should calculate $\Theta^{XY}_{ij}(t)$ and $\Theta^{XY}_{ji}(t)$ separately for a single edge $(i,j)$ when $X\neq Y$.
We obtain the expressions of these asymmetric joint probabilities, i.e., $\Theta^{SI}_{ij}(t)$, $\Theta^{SR}_{ij}(t)$, and $\Theta^{IR}_{ij}(t)$ as follows:
    \begin{eqnarray}    
    \Theta^{SI}_{ij}(t+1)=\Theta^{SS}_{ij}(t)q_{ij}(t)(1-q_{ji}(t)) 
    +\Theta^{SI}_{ij}(t)(1-\lambda)q_{ij}(t)(1-\gamma),
    \end{eqnarray}
    
    \begin{equation}    
    \Theta^{SR}_{ij}(t+1)=    \Theta^{SI}_{ij}(t)(1-\lambda)q_{ij}(t)\gamma+\Theta^{SR}_{ij}(t)q_{ij}(t),
    \end{equation}
and
    \begin{eqnarray}    \label{IR}
    \Theta^{IR}_{ij}(t+1)&=&\Theta^{SR}_{ij}(t)(1-q_{ij}(t))
    +\Theta^{IR}_{ij}(t)(1-\gamma) \nonumber \\
    &+&\Theta^{II}_{ij}(t)(1-\gamma)\gamma  
    +\Theta^{SI}_{ij}(t)(1-(1-\lambda)q_{ij}(t))\gamma.
    \end{eqnarray}
In addition, $q_{ij}(t)$ in Eqs.~(\ref{II})--(\ref{IR}) can be expressed as
    \begin{eqnarray}    
    q_{ij}(t)=\prod_{r=1,r\neq j}^N(1-\lambda A_{ri}h_{ir}(t)),
    \end{eqnarray}
where $h_{ij}(t)=P[\varepsilon_j(t)=I|\varepsilon_i(t)=S]$ stands for the probability that node $j$ is in the I state when node $i$ is in the S state. The conditional probability $h_{ij}(t)$ can be expressed as
    \begin{equation}    
     h_{ij}(t)={\frac{\Theta_{ij}^{SI}(t)}{\Theta_{ij}^{SI}(t)+\Theta_{ij}^{SS}(t)+\Theta_{ij}^{SR}(t)}}.
    \end{equation}
Iterating Eqs.~(\ref{II})--(\ref{IR}) from any meaningful initial condition [e.g.,  $\Theta_{ij}^{SI}(0)=\Theta_{ji}^{SI}(0)=\rho_{0}(1-\rho_{0})$, $\Theta_{ij}^{II}(0)=\rho_{0}^{2}$, $\Theta_{ij}^{SS}(0)=(1-\rho_{0})^{2}$ and $\Theta_{ij}^{RR}(0)=\Theta_{ij}^{SR}(0)=\Theta_{ji}^{SR}(0)=\Theta_{ij}^{IR}(0)=\Theta_{ji}^{IR}(0)=0$] can give the probability of any possible state of edge $(i,j)$ at the final state.  For a network made up of $N$ nodes and $M$ edges, we will have $9M$ equations in total for determining the probabilities of states of all the edges. We refer to the approach  of using the $9M$ equations to solve the SIR model as the SIR--edge--equations (SIRee) approach.
Denote the final value of $\Theta_{ij}^{XY}(t)$ as $\Theta_{ij}^{XY}$.
Then, we can obtain the probabilities of node $i$ in R as    
    \begin{eqnarray}    \label{pr}
    \rho_{i}=\frac1{k_i}\sum_{j=1}^NA_{ij}(\Theta_{ij}^{RR}+\Theta_{ji}^{SR}).
    \end{eqnarray}
Thus, the spreading prevalence can be computed as
    \begin{eqnarray}    \label{rho}
    \rho=\frac1N\sum_{i=1}^N\frac1{k_i}\sum_{j=1}^NA_{ij}(\Theta_{ji}^{RR}+\Theta_{ji}^{SR})
    \end{eqnarray}
Besides, we can get the conditional probability $P(\varepsilon_r=S\vert\varepsilon_j=S)$ in Eq.~(\ref{eq:Iij}) as
    \begin{eqnarray}    \label{psr}
    P(\varepsilon_r=S\vert\varepsilon_j=S)=\frac{\Theta_{jr}^{SS}}{\Theta_{jr}^{SS}+\Theta_{jr}^{SR}}.
    \end{eqnarray}
Define short notations for convenience as follows,
  \begin{eqnarray}    \label{oi}
    o_{i}=(1+\sum_{r=1}^N\lambda A_{ir}\frac{\Theta_{ir}^{SS}}{\Theta_{ir}^{SS}+\Theta_{ir}^{SR}}).
    \end{eqnarray}
Substituting Eqs.~(\ref{pr}) and (\ref{psr}) back into Eq.~(\ref{eq:Iij}) yields the following expression of latent edge influence $\sigma_{ij}$:
  \begin{eqnarray}    \label{importance}
    \sigma_{ij}=\lambda \rho_{i}(1-\rho_{j})o_{j}+\lambda(1-\rho_{i})\rho_{j}o_{i}.
    \end{eqnarray}
Eq. (\ref{importance}) reveals that the influence of each latent edge depends on both the network structure (e.g. the adjacency matrices $A$) and the spreading dynamics (e.g. $\lambda$ and $\gamma$). As described in Sec.~\ref{sec:model},  our strategy for promoting the spreading of the SIR model on networks is based on the addition of the latent edge with highest  influence $\sigma_{ij}$ iteratively.
In order to ensure that we really add the current latent edge with the highest influence, we need to resolve Eqs.~(\ref{II})--(\ref{IR}) and recalculate Eq. (\ref{importance}) after adding any single edge because the network structure changes after each edge addition.

\section{Simulation results}\label{sec:simulation}

This section will present extensive numerical simulations on both synthetic and real--world networks to verify the effectiveness of our approach.
    
\begin{figure}
        \centering
        \includegraphics[width=0.9\textwidth]{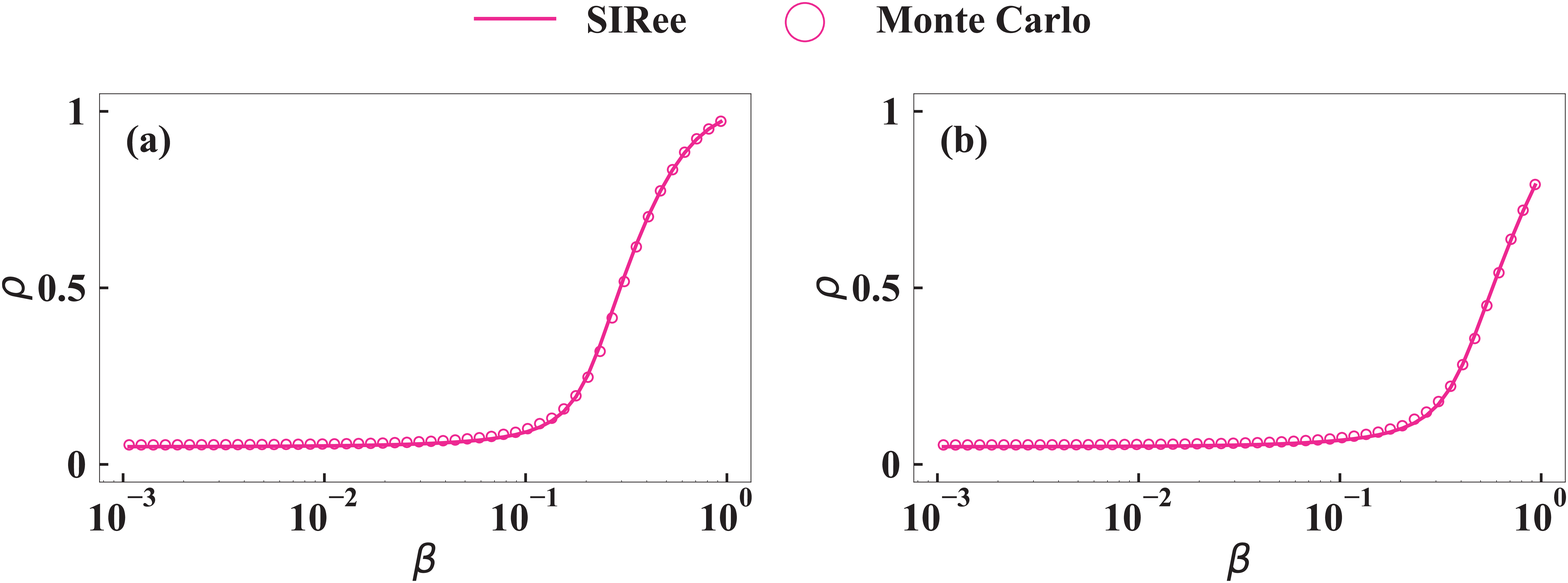}
        \caption{(Color online) Spreading prevalence $\rho$ versus effective transmission probability $\beta$. The spreading prevalence predicted by  Eq.~\ref{rho}  (solid lines) or obtained by Monte Carlo simulations (circles) on the scale--free network with average degree (a) $\left\langle k_{1}\right\rangle=5 $  or (b) $\left\langle k_{2}\right\rangle =3 $. The degree exponents of these two networks are both set to be $\alpha=2.3$. More detailed information of these two synthetic networks can be found in Tab.~\ref{tab:networks}. The recovery probability in the SIR model is set as $\gamma=0.5$.
     }\label{pic1}
    \end{figure}

    To begin with, we test the agreement between our SIR--ee numerical approach proposed in Sec.~\ref{sec:theory} and the empirical simulations for the SIR model.
    Figs.~\ref{pic1} (a) and (b) show the spreading prevalences predicted by Eq.~(\ref{rho}) and obtained by Monte Carlo simulations on two synthetic scale-free (SF) networks $G_{1}$ and $G_{2}$, respectively. These two SF networks have the same degree exponent $\alpha=2.3$ but different average degrees. Specifically, $G_{1}$ has an average degree of $\left\langle k_{1}\right\rangle =5 $ while $G_{2}$ has an average degree of $\left\langle k_{2}\right\rangle =3 $. More information about these two synthetic networks can be found in Tab.~\ref{tab:networks}. As can be seen, there is a marked agreement between the results of our SIR--ee numerical approach and Monte Carlo simulations in the full range of effective transmission probability $\beta$ on both the synthetic network we studied. Thus, it is valid to use our SIR--ee approach to determine the global impact of the SIR model.
    
    Then, we go further to test the performance of our strategy in promoting the spreading of the SIR model on the two synthetic SF networks. 
    As described in Sec.~\ref{sec:theory}, our strategy is to add the latent edge $L$, which has the highest influence $\sigma_{ij}$ calculated by Eq.~(\ref{importance}) iteratively. After the addition of a single edge, we resolve Eqs.~(\ref{II})--(\ref{IR}) and recalculate Eq. (\ref{importance}) to ensure that we really add the current latent edge with the highest influence. 
    For comparison,  we also test three additional strategies. First, we consider the approach to add the latent edge $L^{D}$, which has the largest degree product $f^{d}$, that is, the product of the degree of the nodes connected by the latent edge. This strategy is referred to as the degree--product (DP) strategy in the rest of the paper.
    Similarly, we also consider the strategy to add the latent edge $L^{E}$, which has the largest eigenvector centrality product $f^{e}$, that is, the product of the eigenvector centrality of the nodes connected by the latent edge. We refer to this strategy as the eigenvector--centrality--product (ECP) strategy. Last, we carry out the strategy to add the latent edge $L^{R}$ selected by random and refer to this strategy as the random (RD) strategy.
    Note that we recalculate all the measures in the three strategies after the addition of any single edge, as in the case of our strategy.
    
    \begin{figure}
        \centering
        \includegraphics[width=0.9\textwidth]{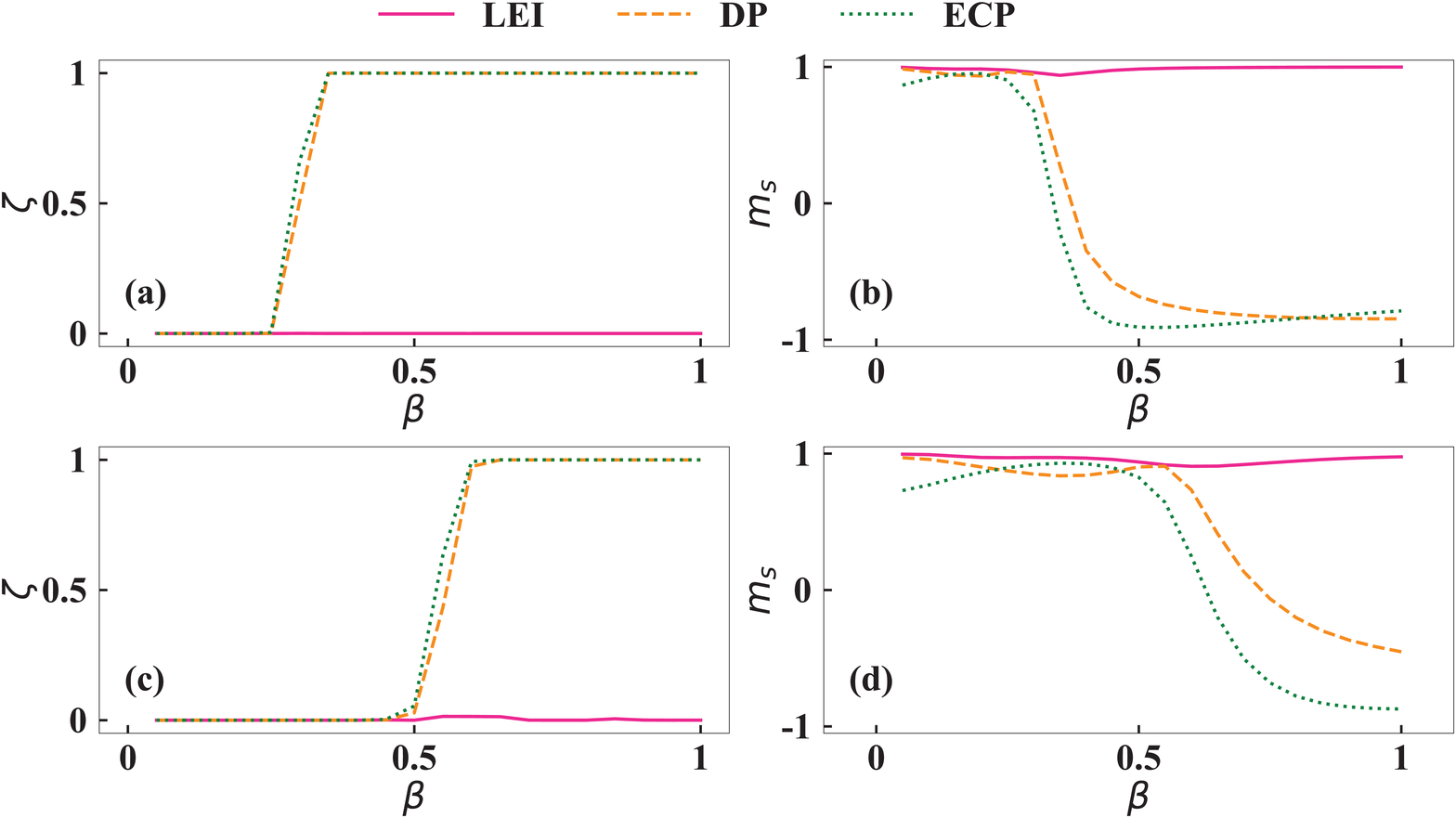}
        \caption{(Color online)  Correlations between the theoretical edge ranks and the numerical edge ranks. The normalized numerical rank $\zeta$ of the optimal latent edge selected by strategy LEI (pink solid line), strategy DP (orange dashed line) or strategy ECP (green dotted line) on the SF network with average degree (a) $\left\langle k_{1}\right\rangle=5 $ or (c) $\left\langle k_{2}\right\rangle =3$. The Spearman’s rank correlation coefficient $m_{s}$ between
the theoretical edge ranks scored by strategy LEI (pink solid line), strategy DP (orange dashed line), or strategy ECP (green dotted line) and the numerical edge ranks on the SF networks with average degree (b) $\left\langle k_{1}\right\rangle=5 $  or (d) $\left\langle k_{2}\right\rangle =3 $. 
        The corresponding degree exponents of both these two synthetic networks are $\alpha=2.3$. More information about these two synthetic networks is presented in Tab.~\ref{tab:networks}. We have set the recovery probability of the SIR model to be $\gamma=0.5$.
    }
        \label{pic2}
    \end{figure}

    Denote $\hat \rho$ as the incremental spreading prevalence obtained by the SIR--ee numerical approach after adding the selected latent edge. Then we rank all the latent edges according to the values of $\hat \rho$. We call this kind of edge rank the numerical edge rank $r$ and denote the normalized numerical edge rank as $\zeta=r/M_{l}$, where $M_{l}$ is the number of all the latent edges. Fig. \ref{pic2} presents the correlations between the theoretical edge ranks scored by different strategies and the numerical edge ranks. Specifically, Figs. \ref{pic2} (a) and (b) demonstrate that the normalized edge rank of the optimal latent edge $L$ selected by our strategy is close to $1/M_{l}$ for the full range of effective transmission probability $\beta$ on both the networks $G_{1}$ and $G_{2}$. The results prove that our strategy performs well in finding the optimal latent edge, which is the key step in promoting strategies. 
    However, the normalized edge ranks of the optimal edges $L^{D}$ and $L^{E}$ become large when $\beta$ is big.  
    Besides, Figs. \ref{pic2} (c) and (d) also show the Spearman rank correlations $m_{s}$ between the theoretical edge ranks scored by different strategies and the numerical edge ranks, that is, 
    \begin{equation}
    m_{s}=1-6\frac{\sum^{M_{l}}_{l=1}(r_{l}-\hat r_{l})^{2}}{M_{l}(M_{l}^{2}-1)}，
    \end{equation}
where $r_{l}$ and $\hat r_{l}$ denote the theoretical edge rank and numerical edge rank of edge $l$, respectively.   
It can be seen that the Spearman rank correlation between the theoretical edge ranks scored by our strategy and the numerical edge ranks is close to $1$ for the full range of effective transmission probability $\beta$ on both networks. This suggests that our strategy can well predict the overall numerical ranks of the latent edges. 
However, the Spearman correlation between the theoretical edge ranks scored by the strategy DP or ECP, and the numerical edge ranks are close to $1$ only for $\beta$ of small values. 
This can be explained by the fact that nodes with a high degree or eigenvector centrality will be infected or informed with a larger probability compared with those nodes with small centralities when $\beta$ is small. If we add the latent edges between them, then these high--centrality nodes together with their neighbors can form an infected or informed cluster that facilitates the spreading. Thus the DP and ECP strategies perform well in finding the optimal latent edge or predicting the overall numerical ranks when $\beta$ is small. However, when $\beta$ becomes large, the globally spreading outbreak occurs; thus, connecting the nodes with high centralities becomes unnecessary, but additional connections to those nodes with low centrality are required for the promoting of the spreading. Therefore, both the DP and ECP strategies fail.
Note that random strategy is useless in finding the optimal latent edge or predicting the numerical ranks of the latent edges; thus, we have not included the corresponding results of random strategy here.
All in all, Fig.~\ref{pic2} shows strong evidence for the potential superiority of our strategy in promoting the spreading of the SIR model. 

\begin{figure}
        \centering
        \includegraphics[width=0.9\textwidth]{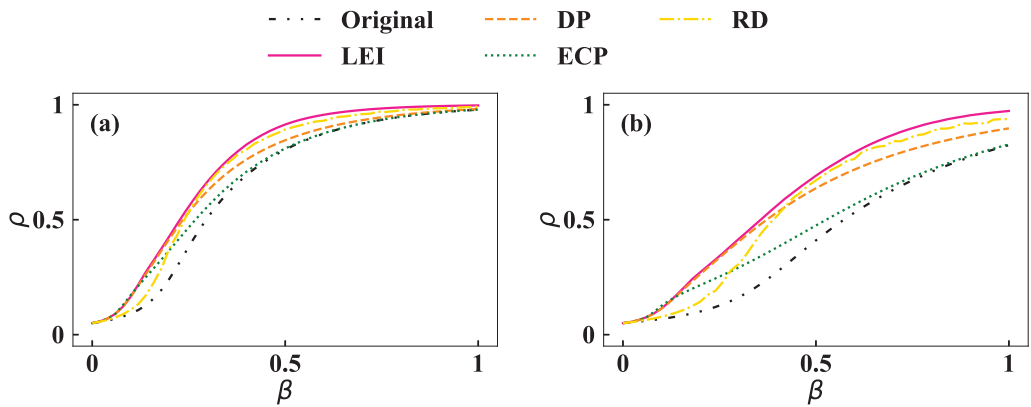}
        \caption{(Color online) Performance of different strategies versus effective transmission probability $\beta$. The original spreading prevalence on the SF network (black dash--dot--dot line)  with average degree (a) $\left\langle k_{1}\right\rangle=5 $ or (b) $\left\langle k_{2}\right\rangle=3 $.  The corresponding spreading prevalences after adding a number of $N/2$ edges using strategy LEI, DP, ECP and RD are denoted by pink solid line,  orange dashed line, green dotted line and yellow dash--dot line, respectively.  The degree exponents of both these two synthetic networks are $\alpha=2.3$ and the recovery probability of the SIR model is $\gamma=0.5$. Tab.~\ref{tab:networks} shows the detailed information of these two synthetic networks.      }\label{pic3}
    \end{figure}

\begin{figure}
        \centering
        \includegraphics[width=0.9\textwidth]{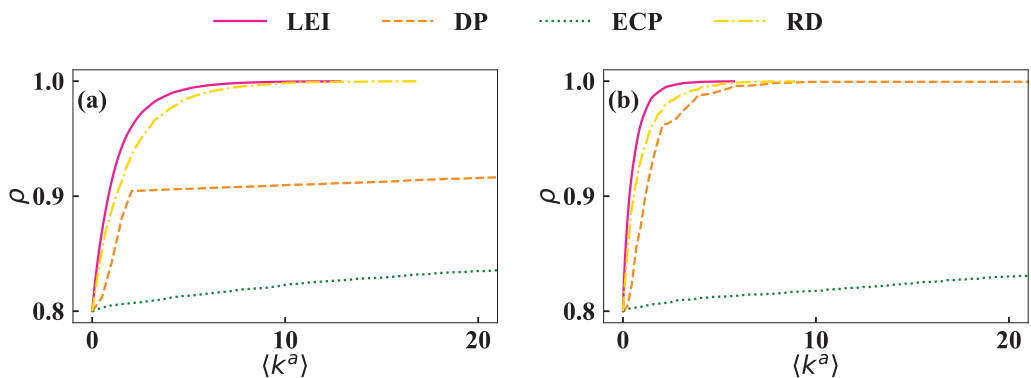}
        \caption{(Color online) Spreading prevalence $\rho$ versus incremental average degree $\left\langle k^{a}\right\rangle$. The spreading prevalence as a function of the incremental average degree on the SF networks with original average degree (a) $\left\langle k_{1}\right\rangle=5 $ or (b) $\left\langle k_{2}\right\rangle=3 $. We compare the results of strategy LEI (pink solid line), strategy DP (orange dashed line), strategy ECP (green dotted line) and strategy RD (yellow dash--dot line). The recovery probabilities of the SIR model on the two networks are both set to be $\gamma=0.5$. Besides, we choose the transmission probabilities $\lambda$ such that the original spreading prevalences of the SIR model are about $\rho=0.8$ for both the networks, i.e., $\lambda=0.252$ and $\lambda=0.487$ for the network with original  average degree $\left\langle k_{1}\right\rangle=5 $ and $\left\langle k_{2}\right\rangle=3$, respectively.
The detailed information of the two synthetic SF networks is shown in Tab. (\ref{tab:networks}).
     }\label{pic4}
    \end{figure}

\begin{figure}
        \centering
        \includegraphics[width=0.9\textwidth]{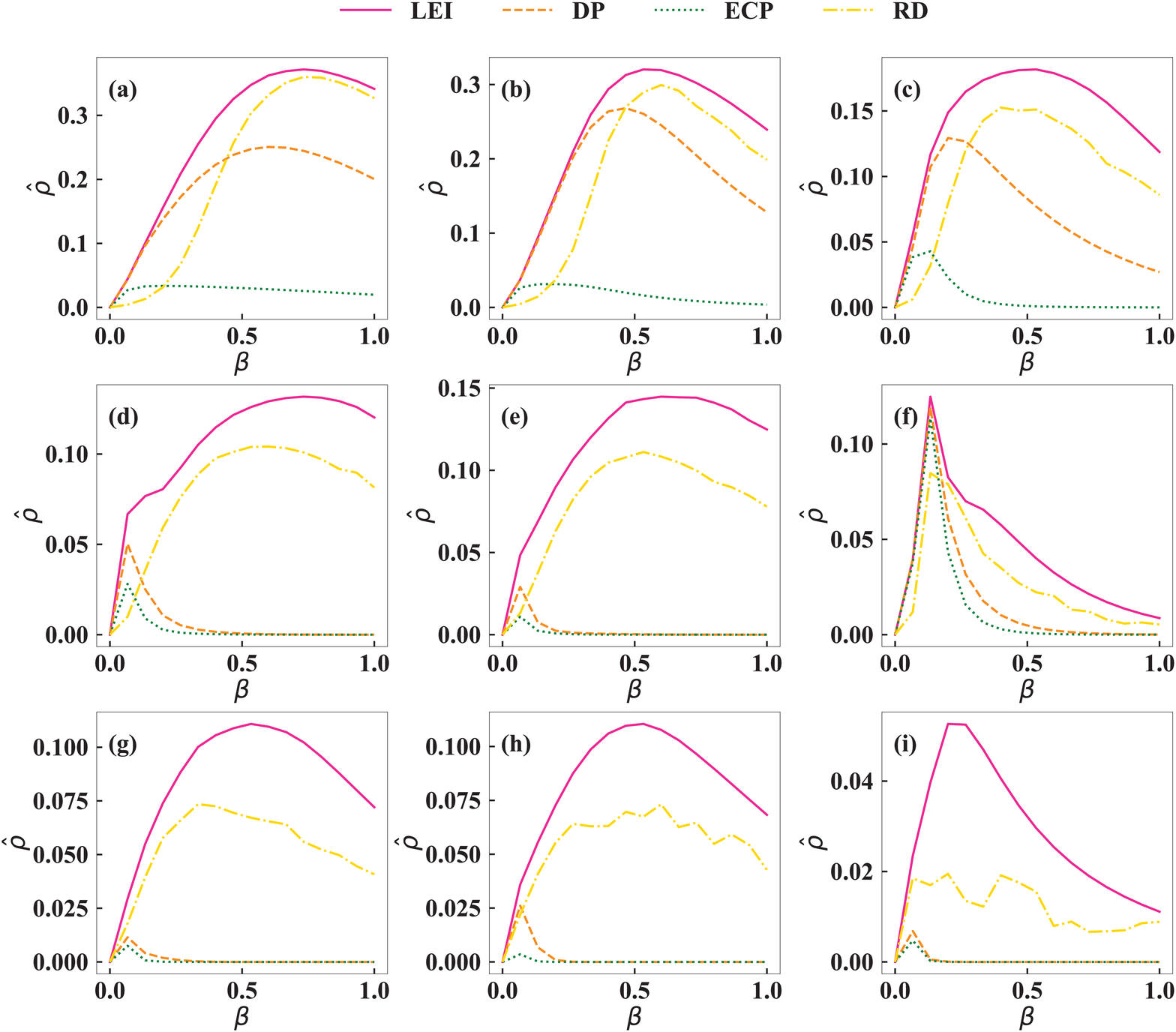}
        \caption{(Color online) Incremental spreading prevalence $\hat \rho$ versus effective transmission probability $\beta$. The incremental spreading prevalence after adding a number of $N/2$ edges by strategy LEI (pink solid line) , strategy DP (orange dashed line), strategy ECP (green dotted line) or strategy RD (yellow dash--dot line) as a function of the effective transmission probability on the real--world network (a) ca--CSphd, (b) 1138--bus,
 (c) Air traffic control, (d) web--EPA, (e) tech--routers--rf , (f) Physicians, (g) inf--USAir97, (h) econ--wm1, or (i) Jazz musicians. Detailed information of these real--world networks is presented in Tab. (\ref{tab:networks}) and the recovery probability in the SIR model is set as $\gamma=0.5$.
     }\label{pic5}
    \end{figure}

Afterward, Figs.~\ref{pic3} and \ref{pic4} give intuitive demonstrations of the performance of different strategies on the two synthetic networks from two perspectives. 
On the one hand, Fig.~\ref{pic3} compares the original spreading prevalence and the spreading prevalence after adding a number of $N/2$ edges (that is, increasing the average degree of the network by $1$) using different strategies. The results lead to the conclusion that our strategy performs the best in promoting the spreading of the SIR model for the full range of the effective transmission probability $\beta$ on both networks. Meanwhile, the DP and ECP strategies have good performance only when $\beta$ is small, and the RD strategy performs well only for $\beta$ of large values. It also should be mentioned that the incremental spreading prevalences are much larger in the more sparse network $G_{2}$ after adding the same number of edges by our strategy. That is to say, the effectiveness of our strategy is more obvious in sparse networks, which are common in the real world.
On the other hand,  Fig.~\ref{pic4} demonstrates that our strategy can bring the fastest full--blown break--out of the SIR model. In the numerical simulations, we set the recovery probability to be $\gamma=0.5$ and choose the transmission probability $\lambda$ such that the original spreading prevalence of the SIR model is about $\rho=0.8$ for both the two synthetic networks, that is, $\lambda=0.252$ and $\lambda=0.487$ for $G_{1}$ and $G_{2}$, respectively. It can be observed that our strategy performs the best in increasing the spreading prevalence to $\rho=1$ on both networks. Besides, the DP and ECP strategy both perform worse than the RD strategy since the value of effective transmission probabilities $\beta$ are relatively large on both networks. These results about the three strategies (i.e., DP strategy, ECP strategy, and RD strategy) coincide with the findings we obtained from Fig.~\ref{pic3}. 
Sum up, Figs.~\ref{pic3} and \ref{pic4} give the direct proofs of the effectiveness and superiority of our strategy.

Finally, we test our strategy on $9$ real--world networks: (a) ca—CSphd~\cite{nr}; (b) 1138—bus~\cite{nr};
 (c) Air traffic control~\cite{10.1145/2487788.2488173}; (d) web-EPA~\cite{nr}; (e) tech-routers-rf~\cite{nr} ; (f) Physicians~\cite{10.1145/2487788.2488173}; (g) inf-USAir97~\cite{nr}; (h) econ-wm1~\cite{nr}; and (i) Jazz musicians~\cite{10.1145/2487788.2488173}. Detailed information of these real--world networks is presented in Tab. \ref{tab:networks}. They cover a wide range of average degree (between 2.035 and 27.697). We plot the incremental spreading prevalence $\hat \rho$ after increasing the average degree by $1$ (that is, adding a number of $N/2$ edges) as a function of the effective transmission probability $\beta$ in Fig.~\ref{pic5}. 
 It can be seen that our strategy leads to the largest incremental spreading prevalence $\hat \rho$ for the full range of effective transmission probability $\beta$ on all the $9$ real--world networks. Besides, the DP and ECP strategies perform better than the random strategy only for $\beta$ of small values. Moreover, the incremental spreading prevalence $\hat \rho$ is larger in the network with a smaller average degree.
 The results of these real--world networks are in concordance with the conclusions we draw on the synthetic networks $G_{1}$ and $G_{2}$.

        \begin{table}
        \centering
        \caption{Basic statistics of the two synthetic networks and nine real--world networks employed in this study: the number of
            nodes $N$, the number of edges $M$, the maximum degree $k_{max}$, the average degree $\left\langle k\right\rangle $, and the second moment of the degree distribution $\left\langle k^{2}\right\rangle$.}
        \begin{tabular}{llllllllll}
            \br
            Name                    &$N$ & $M$ & $k_{max}$& $\left\langle k\right\rangle $  & $\left\langle k^{2}\right\rangle$     \\  \br
            SF2.3                   & 200  & 500 &14     & 5                                            & 31.27 \\
            sparse SF2.3            & 200  & 500 &9    & 3                                        & 11.92 \\
             ca--CSphd       & 1025  & 1043 &46    & 2.035                                        & 12.166 \\
              1138--bus       & 1038  & 1458 &17    & 2.562                                       & 9.814 \\
             Air traffic control     & 1226 & 2408 &34    & 3.928                                             & 28.899 \\
              web-EPA    & 4253 & 8897 &175     & 4.184                                             & 118.451 \\
              tech-routers-rf     & 2113 & 6632 &109     & 6.277                                             & 135.704 \\
              Physicians              & 117  & 465 &26    & 7.95                                              & 79.162 \\
              inf-USAir97       & 332  & 2126  &139   & 12.807                                       & 568.163 \\
              econ-wm1       & 258  & 2389  &106   & 18.519                                       & 917.434 \\
            Jazz musicians          & 198  & 2742 &100    & 27.697                                          &1070.242 \\
              \br
                    \end{tabular}
        \label{tab:networks}
    \end{table}

    \section{Conclusions}\label{sec:conclusion}
    Promoting some typical spreading dynamics (for instance, the spreading of information, vaccination guidance, commercial message, innovation, and political movements) in networked systems can be of both theoretical and practical importance. In this study, we proposed an effective edge--based strategy for promoting the spreading dynamics of the SIR model on complex networks. 
    
 To be specific, we first quantified the potential influence that the addition of each latent edge could cause to the spreading dynamics by a mathematical model. This mathematical model could also facilitate the determination of the spreading prevalence. Then, we strategically added the latent edges to the original networks according to the potential influence of each latent edge. Note that previous approaches for promoting the spreading dynamics on complex networks mostly only consider either the structure of networks or spreading dynamics. However, our strategy incorporates both the information of network structure and spreading dynamics. 
Extensive numerical simulations verified the effectiveness of our strategy and demonstrated that our strategy outperforms those static approaches, such as adding the latent edge between nodes with the highest degree or eigenvector centrality.
  
 This study provides an effective approach for promoting the spreading of the SIR model by modifying the network structure slightly and helps to understand what a better network structure for the spreading dynamics is. 
Besides, the theoretical framework we developed in this study offers inspirations for further investigations on edge--based promoting strategies for other spreading models.
    
    \section*{Acknowledgements}
    
    This work was partially supported by the China Postdoctoral Science Special Foundation (Grant No.~2019T120829), National Natural Science Foundation of China (Grant Nos.~61903266), Fundamental Research Funds for the Central Universities, and Sichuan Science and Technology Program (NO. 20YYJC4001).

\section*{References}
\providecommand{\newblock}{}

%%\bibliography{xianjiajun}

\end{document}